\documentclass[
]{ceurart}

\usepackage{scolia26-sr4cs-frame}

\begin{document}
\copyrightyear{2024}
\copyrightclause{Copyright for this paper by its authors. Use permitted under Creative Commons License Attribution 4.0 International (CC BY 4.0).}

\conference{SCOLIA 2026, the Second International Workshop on Scholarly Information Access, ECIR 2026, 2nd April 2026, Delft, Netherland}

\title{A Collection of Systematic Reviews in Computer Science}

\author[1]{Pierre Achkar}[orcid=0009-0007-0791-9078]
\author[2]{Tim Gollub}[orcid=0000-0003-1737-6517]
\author[3]{Martin Potthast}[orcid=0000-0003-2451-0665]

\address[1]{Leipzig University, Fraunhofer ISI Leipzig}
\address[2]{Bauhaus-Universität Weimar}
\address[3]{Kassel University, hessian.AI, ScaDS.AI}

\begin{abstract}
Systematic reviews are the standard method for synthesizing scientific evidence, but their creation requires substantial manual effort, particularly during retrieval and screening. While recent work has explored automating these steps, evaluation resources remain largely confined to the biomedical domain, limiting reproducible experimentation in other domains. This paper introduces SR4CS, a large-scale collection of systematic reviews in computer science, designed to support reproducible research on Boolean query generation, retrieval, and screening. The corpus comprises \num{1212}~systematic reviews with their original expert-designed Boolean search queries, \num{104316}~resolved references, and structured methodological metadata. For controlled evaluation, the original Boolean queries are additionally provided in a normalized, approximated form operating over titles and abstracts. To illustrate the intended use of the collection, baseline experiments compare the approximated expert Boolean queries with zero-shot LLM-generated Boolean queries, BM25, and dense retrieval under a unified evaluation setting. The results highlight systematic differences in precision, recall, and ranking behavior across retrieval paradigms and expose limitations of naive zero-shot Boolean generation. SR4CS is released under an open license on Zenodo,
\footnote{\url{https://doi.org/10.5281/zenodo.17163932}}
together with documentation and code,%
\footnote{\url{https://github.com/webis-de/scolia26-sr4cs}}
to enable reproducible evaluation and future research on scaling systematic review automation.
\end{abstract}


\maketitle

\section{Introduction}

Systematic reviews are the standard method for synthesizing evidence on focused research questions. They are valued for their completeness, transparency, and reproducibility, and their results often serve as a guide for future research, policy, and practice~\cite{Liberati2009ThePSA,Lam2019SystematicLRA}. Conducting a systematic review is time-consuming and involves steps such as defining research questions, retrieving candidate papers, screening for relevance, and synthesizing findings~\cite{CochraneCh4}. Retrieving candidate papers is a vital step, typically accomplished by translating the review topic into complex Boolean queries to support interpretability and reproducibility. These queries define the pool of eligible candidate studies and directly influence the outcome of the review because any loss in recall during retrieval can hardly be recovered~\cite{MACFARLANE2022200091}.

In recent years, increasing efforts have been made to automate systematic reviews to reduce the cost of their creation, both end-to-end and in individual phases such as query formulation, screening, and synthesis. Work on automatic Boolean query generation ranges from computational adaptations of expert-designed strategies~\cite{DBLP:journals/ir/ScellsZK21} to recent investigations using large language models~(LLMs), which show potential but also exhibit substantial trade-offs in precision and recall~\cite{DBLP:conf/sigir/WangSKZ23, DBLP:conf/ecir/WangSZPKZ24}. LLMs have also been explored for screening, where recall-oriented calibration can yield promising zero-shot performance~\cite{DBLP:conf/ecir/WangSZPKZ24}, and agent-based systems are emerging to orchestrate multiple stages of the systematic review workflow~\cite{DBLP:journals/corr/abs-2403-08399}. Despite this progress, the reliability and reproducibility of automated methods remain open research questions that require controlled evaluation.

However, evaluation resources for systematic review automation remain heavily concentrated in the biomedical domain. Datasets such as the SIGIR~2017 SysRev Query Collection~\cite{scells2017sigir}, the CLEF TAR~2019 corpus~\cite{DBLP:conf/clef/KanoulasLAS19}, and the Seed Studies collection~\cite{DBLP:conf/sigir/WangSCKZ22} provide valuable test collections for medical systematic reviews, but there is no comparable large-scale resource for other domains. In the absence of domain-specific benchmarks, evaluating retrieval and screening methods often requires costly expert involvement, constraining scale and reproducibility.

\begin{table*}[t]
\centering
\caption{Benchmark datasets for systematic review retrieval.}
\begin{tabularx}{\textwidth}{@{}llrXX@{}}
\toprule
\bfseries Dataset & \bfseries Domain & \bfseries Topics &
\bfseries Reference pool & \bfseries Tasks \\
\midrule
SIGIR 2017 SysRev\footnotemark
& Biomedicine & 94
& 26M MEDLINE records
& Retrieval; screening \\

CLEF TAR 2019\footnotemark
& Biomedicine & 123
& PubMed baseline (2019)
& Retrieval; screening \\

Seed-Studies~2022\footnotemark
& Biomedicine & 40
& PubMed retrieved sets + snowballing
& Retrieval (seed-based); screening; citation chasing \\
\midrule
SR4CS
& Computer science & 1,212
& 104k references (89k with abstracts)
& Retrieval; screening \\
\bottomrule
\end{tabularx}
\label{table-sr-datasets}
\end{table*}

To address this gap, we introduce SR4CS, the first large-scale test collection of systematic reviews in computer science, paired with the original Boolean search strategies reported by the authors and curated reference pools. Each review includes the original Boolean query, alongside an approximation normalized to a unified title-and-abstract-only retrieval setting for controlled evaluation. In addition, each review is associated with structured methodological metadata, including research objectives, inclusion and exclusion criteria, databases searched, and temporal constraints. In total, SR4CS contains \num{1212}~systematic reviews with \num{104316}~resolved references, of which \num{89447}~include abstracts. The dataset is publicly available together with documentation and code, enabling reproducible evaluation of Boolean query translation, retrieval effectiveness, and screening methods in computer science.
\section{Related Work}

Existing evaluation resources for retrieval and screening in systematic reviews have largely focused on the biomedical domain. While these resources vary in scale, design, and supported tasks, they all provide reusable test collections for evaluating retrieval effectiveness, screening, and related automation methods. Table~\ref{table-sr-datasets} summarizes the most relevant datasets and contrasts them with~SR4CS.

\setcounter{footnote}{3}
\footnotetext{\url{https://github.com/ielab/SIGIR2017-SysRev-Collection}}
\setcounter{footnote}{4}
\footnotetext{\url{https://github.com/ielab/sysrev-seed-collection}}
\setcounter{footnote}{5}
\footnotetext{\url{https://github.com/CLEF-TAR/tar}}

The SIGIR~2017 SysRev Query Collection~\cite{scells2017sigir} is based on $\sim$26~million MEDLINE records derived from 94~Cochrane reviews published between~2014 and~2016. The reported search strategies were converted into executable queries, and included and excluded references were mapped to MEDLINE identifiers, enabling controlled experiments on retrieval and screening prioritization in the biomedical domain.

The Seed Studies Collection~\cite{DBLP:conf/sigir/WangSCKZ22} was developed to support research on seed-driven search strategies. It comprises 40~medical systematic review topics curated by information specialists and includes the original PubMed Boolean queries, seed studies, retrieved records, and final included references. By explicitly distinguishing genuine seed studies from pseudo-seeds, the collection enables more realistic evaluation of automatic query formulation, screening prioritization, and citation chasing.

The CLEF TAR~2019 corpus~\cite{DBLP:conf/clef/KanoulasLAS19}, developed as part of the CLEF lab on technology-assisted reviews, is also grounded in Cochrane reviews and PubMed. It includes expert queries, retrieved records, and relevance judgments for 123~topics across several medical review types, and supports evaluation of protocol querying and screening prioritization.

SR4CS extends these efforts beyond biomedicine by providing the first large-scale test collection of systematic reviews in computer science. It preserves the original Boolean search strategies as reported in the reviews, while also providing normalized, approximated variants for controlled evaluation. With \num{1212}~systematic reviews and resolved reference pools, SR4CS supports reproducible evaluation of Boolean query translation, retrieval effectiveness across paradigms, and screening methods in a domain where comparable evaluation resources have previously been lacking.

\section{The SR4CS Corpus}

\begin{figure*}[htbp]
\centering
\includegraphics[width=\textwidth]{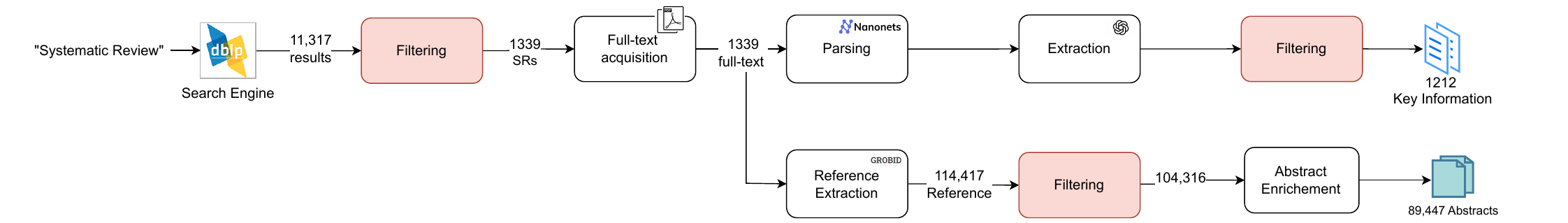}
\caption{SR4CS-25 construction pipeline: collection, filtering, parsing, extraction, and reference resolution.}
\label{fig:pipeline}
\end{figure*}

The corpus construction comprises three main stages: data collection and filtering, parsing and information extraction, and resolving references. Figure~\ref{fig:pipeline} provides an overview of the pipeline.

\subsection{Data Collection}

A set of candidate systematic reviews was retrieved 
from DBLP by searching titles for ``systematic review'', yielding \num{11317}~results. We applied filter rules to retain only genuine systematic reviews that required clear wording (e.g., ``systematic review of/on''), peer review procedures, open access, and valid~DOIs. Outliers due to extreme page counts were removed. After filtering, \num{1339}~candidate reviews remained. In the following steps, only reviews that explicitly stated their Boolean queries were considered for inclusion in the final dataset.

\subsection{Parsing and Extraction}

The PDF files of the systematic reviews were downloaded and converted into Markdown format for further processing. To achieve the conversion, \texttt{Nanonets-OCR-s},%
\footnote{\url{https://huggingface.co/nanonets/Nanonets-OCR-s}}
a modern OCR model that produces semantically enriched Markdown text rather than plain text, was used. Unlike traditional OCR tools, it can recognize LaTeX equations, describe figures and tables, and preserve document structure. This functionality was particularly important because many reviews contain Boolean queries and other important information in figures or tables.
 
The documents were then processed in a zero-shot settings using GPT-4.1 Mini to extract a series of structured fields:
(i)~databases used,
(ii)~Boolean queries reported,
(iii)~year range,
(iv)~language restrictions,
(v)~inclusion and exclusion criteria,
(vi)~main topic,
(vii)~objective,
(viii)~research questions, and
(ix)~whether snowballing or citation chasing was performed.
This approach follows recent work demonstrating the feasibility of LLM-based zero-shot data extraction in scientific documents~\cite{DBLP:journals/corr/abs-2303-05352,gartlehner24extract}.

To assess extraction quality, the extracted fields were manually inspected for errors and inconsistencies. Additionally, we randomly selected~10\% of the reviews and manually extracted the same information independently. In~85\% of the sample cases, the fields extracted by the LLM matched the manual extraction at the field level, and the errors found were minimal, such as minor discrepancies in the wording of the inclusion/exclusion criteria or the specification of certain field values (e.g., language restrictions) based on indirect clues rather than explicit statements. These findings indicate that zero-shot LLM-based extraction provides a sufficiently reliable basis for large-scale corpus construction, while still leaving room for more robust, feedback-driven extraction methods in future extensions. Finally, we excluded reviews that did not explicitly state their Boolean queries, resulting in a final dataset of \num{1212}~systematic reviews.

\subsection{Reference Resolution}

References were extracted from each systematic review via a hybrid pipeline combining AnyStyle%
\footnote{\url{https://github.com/inukshuk/anystyle}} 
and GROBID,%
\footnote{\url{https://github.com/kermitt2/grobid}}
a configuration proven to be highly effective for structured citation analysis~\cite{DBLP:journals/jodl/BackesISM24}. First, non-scientific items such as reports, interviews, and patents were removed, excluding \num{10101}~entries. The remaining references were enriched with abstracts from public bibliographic APIs, including Crossref, OpenAlex, Semantic Scholar, PubMed, Europe PMC, and the arXiv~API. During resolution, DOI matches were prioritized, while references without a DOI were handled through a high-threshold title-author match and consistency checks for the publication year to minimize false positives. Overall, this process produced a consolidated reference pool on which the SR4CS corpus is based.

\subsection{Corpus Statistics}

Table~\ref{tab:corpus-stats} summarizes the size and structural properties of the SR4CS corpus, providing context for the retrieval and screening experiments, while Table~\ref{tab:db-stats} reports the bibliographic databases most frequently cited in the reviewed systematic reviews, reflecting common data sources used in computer science literature reviews.

\begin{table}[t]
\centering
\caption{Core statistics of the SR4CS corpus.}
\small
\begin{tabular}{lr}
\hline
\textbf{Statistic} & \textbf{Value} \\
\hline
Systematic reviews & 1,212 \\
Total resolved references & 104,316 \\
References with abstracts & 89,447 \\
Average references per review & $\sim$83 \\
Median references per review & 74 \\
Minimum references per review & 4 \\
Maximum references per review & 459 \\
\hline
\end{tabular}
\label{tab:corpus-stats}
\end{table}

\begin{table}[t]
\centering
\caption{Most frequently used bibliographic databases in SR4CS.}
\small
\begin{tabular}{lr}
\hline
\textbf{Database} & \textbf{Count} \\
\hline
Scopus & 649 \\
ScienceDirect & 478 \\
Web of Science & 418 \\
ACM Digital Library & 439 \\
IEEE Xplore & 387 \\
Google Scholar & 346 \\
PubMed & 150 \\
\hline
\end{tabular}
\label{tab:db-stats}
\end{table}

\section{Experiments and Intended Use}

The SR4CS corpus is released to support reproducible research on retrieval and screening for systematic reviews in computer science. The dataset is designed for three main tasks:
(i)~evaluating automatic Boolean query generation against approximations of expert-designed Boolean queries,
(ii)~comparing retrieval paradigms under realistic systematic review topics, and
(iii)~investigating screening behavior using curated reference pools.
To illustrate its use, we present baseline experiments for Tasks~(i) and~(ii), evaluating Boolean, probabilistic (BM25), and dense retrieval methods using titles and abstracts only.

\subsection{Approximating Expert-Designed Boolean Queries}

A Boolean retrieval baseline is constructed by approximating the expert-designed queries reported in the original reviews. Since these queries rely on database-specific syntax and metadata fields, normalization is required to obtain a unified Boolean formulation operating over document titles and abstracts.

\textbf{Query rewriting.}
Reported Boolean queries are converted into valid SQLite FTS5 \texttt{MATCH} syntax using \texttt{GPT-4.1-mini} in a zero-shot setting. The query structure is preserved, restricted to the \texttt{title} and \texttt{abstract} fields, and stripped of unsupported metadata filters. A random 25\%~sample is manually verified, with only seven queries requiring minor corrections, primarily for verbatim phrases.

\textbf{Index and execution.}
A SQLite FTS5 index is created using the \texttt{unicode61} tokenizer with diacritics folded and prefix indexing (2--10) to account for common spelling and formatting variations. All refined queries are executed against this index; when multiple Boolean queries are associated with a review, their result sets are merged via union.

\subsection{Alternative Retrieval Baselines}

Using the Boolean approximation as a reference point, we evaluate three additional baselines operating on the same corpus and fields: (i)~zero-shot Boolean queries generated from the review title and objective using \texttt{GPT-4.1-mini},
(ii)~a keyword-based BM25 baseline, and
(iii)~a dense semantic retriever based on the \texttt{all-MiniLM-L6-v2} Sentence-BERT model.

For both BM25 and dense retrieval, queries are constructed from the review title, the objective, and their concatenation. Results are reported for the best-performing variant (title + objective). Retrieval is capped at \num{1000}~documents per query to reflect a fixed screening budget and ensure comparability across retrieval methods.

\subsection{Evaluation and Results}

All retrieval outputs are compared against the curated reference pools to compute precision, recall, F\textsubscript{1}, and F\textsubscript{3}, macro-averaged across reviews. For ranked retrieval methods, we additionally report MAP, P@10, and~R@100.

Although Boolean retrieval is inherently unranked, Boolean result sets are deterministically ranked using SQLite’s internal BM25 scoring function for the purpose of ranking-based evaluation.

\begin{table}[t]
\centering
\caption{Screening effectiveness of approximated expert-designed and automated retrieval approaches. Non-Boolean methods are capped at the top 1{,}000 retrieved documents.}
\label{table-evaluation-unrankend}
\begin{tabular}{@{}lcccc@{}}
\toprule
\bfseries Method & \bfseries Precision & \bfseries Recall & \bfseries F\textsubscript{1} & \bfseries F\textsubscript{3} \\
\midrule
Expert Boolean (Approx.) & \bfseries 0.352 & 0.342 & \bfseries 0.224 & 0.241 \\
GPT4.1-Mini Boolean (ZS) & 0.298 & 0.099 & 0.095 & 0.085 \\
BM25 (Title+Obj, Top-1k) & 0.041 & 0.512 & 0.074 & 0.223 \\
Dense (Title+Obj, MiniLM) & 0.057 & \bfseries 0.702 & 0.104 & \bfseries 0.309 \\
\bottomrule
\end{tabular}
\end{table}
\begin{table}[t]
\centering
\caption{Ranking effectiveness of approximated expert-designed and automated retrieval approaches on SR4CS-25. Boolean result sets are deterministically ordered using BM25 for ranking-based evaluation.}
\label{table-evaluation-rankend}
\begin{tabular}{@{}lccc@{}}
\toprule
\bfseries Method & \bfseries MAP & \bfseries P@10 & \bfseries R@100 \\
\midrule
Expert Boolean (Approx.) & 0.173 & 0.490 & 0.238 \\
GPT4.1-Mini Boolean (ZS) & 0.054 & 0.256 & 0.083 \\
BM25 (Title+Obj) & 0.175 & 0.432 & 0.262 \\
Dense (Title+Obj) & \bfseries 0.281 & \bfseries 0.545 & \bfseries 0.373 \\
\bottomrule
\end{tabular}
\end{table}

Table~\ref{table-evaluation-unrankend} reports set-based screening effectiveness, while Table~\ref{table-evaluation-rankend} reports ranking effectiveness across all evaluated methods. Together, these results provide a consistent and reproducible reference point for evaluating Boolean query generation and alternative retrieval paradigms within the SR4CS corpus.
\section{Discussion}

The development of SR4CS builds on recent advances in document processing and natural language technologies for large-scale extraction from scientific publications. OCR parsers based on vision–language models improve handling of complex PDF layouts~\cite{olmocr}, LLMs support the extraction of methodological information from scientific texts~\cite{gartlehner24extract, DBLP:journals/npjdm/LaiLBLPLHZXTCZELLSSSBYH25}, and GROBID continues to perform strongly in reference extraction~\cite{DBLP:journals/jodl/BackesISM24}. By integrating these components into a unified pipeline, SR4CS demonstrates a reproducible approach to constructing systematic review corpora and provides a transferable methodology for building evaluation resources across scientific domains.

The retrieval results clarify how different paradigms behave when evaluated under the same title-and-abstract-only constraints imposed during corpus construction. Approximated versions of expert-designed Boolean queries achieve the highest precision~(0.352), confirming that Boolean logic remains the most effective means of controlling screening effort; the moderate recall~(0.342) primarily reflects the approximation process, which strips database-specific operators, controlled vocabularies, and citation-based expansion techniques rather than shortcomings of the original strategies. Dense retrieval, in contrast, delivers substantially higher recall~(0.702) and the strongest ranking effectiveness (MAP~0.281, P@10~0.545), demonstrating robustness to vocabulary mismatch and terminological variation in metadata-constrained settings. The zero-shot GPT-4.1- Mini Boolean baseline performs poorly across all metrics (recall~0.099), providing a clear negative result: naive zero-shot Boolean generation with an off-the-shelf model fails to recover either the coverage or selectivity of expert-designed queries, indicating the need for more structured and adaptive query construction approaches. Overall, these results establish SR4CS as a targeted diagnostic benchmark for exposing precision--recall trade-offs and failure modes in retrieval and automated query generation.

Several directions for future extensions of SR4CS are possible. First, deeper integration with open scholarly infrastructures such as OpenAlex would support reference coverage, metadata enrichment, and query execution within an open and reproducible ecosystem, enabling end-to-end studies that rely on open identifiers and transparent metadata. Second, extending the corpus-construction workflow to additional domains beyond computer science (e.g., education and social science) would broaden the benchmark's scope and help assess how retrieval and query-generation methods transfer across differences in terminology and reporting conventions. Third, future work could explore agentic LLM-based information extraction frameworks that use iterative feedback and explicit grounding in source text to improve extraction fidelity and consistency~\cite{Li2025.09.11.675507,DBLP:journals/corr/abs-2510-00276}; such approaches may also make it easier to attach verifiable evidence to extracted fields and to diagnose error modes in downstream evaluation.

 \section{Conclusion}

We introduced SR4CS, a large-scale test collection of systematic reviews in computer science, designed to support reproducible research on Boolean query generation, retrieval, and screening. By providing approximated versions of expert-designed Boolean queries, curated reference pools, and a controlled evaluation setting over titles and abstracts, SR4CS enables systematic comparisons of retrieval paradigms and query formulation strategies beyond the biomedical domain. The dataset is released with documentation and code to facilitate reuse, benchmarking, and future work on systematic review automation, retrieval evaluation, and query generation in computer science and related fields.

{\fontsize{9.4}{11}\selectfont
\bibliography{scolia26-sr4cs-lit}
}
\end{document}